\documentclass[10pt, conference, letterpaper]{IEEEtran}
\usepackage[letterpaper, left=0.65in, right=0.65in, bottom=1.03in, top=0.75in]{geometry}
\IEEEoverridecommandlockouts
\usepackage{cite}
\usepackage{amsmath,amssymb,amsfonts}
\usepackage{graphicx}
\usepackage[caption=false,font=footnotesize]{subfig}
\usepackage{textcomp}
\usepackage{xcolor}
\usepackage{tabularx}
\usepackage{multirow}
\usepackage{diagbox}
\usepackage[binary-units]{siunitx} 
\DeclareSIUnit{\bps}{bps}
\usepackage{amsthm}
\def \tr {\mathrm{tr}}

\begin{document}

\title{A Heterogeneous Massive MIMO Technique for Uniform Service in Cellular Networks}

\author{\IEEEauthorblockN{Wei Jiang\IEEEauthorrefmark{1} and Hans D. Schotten\IEEEauthorrefmark{2}}
\IEEEauthorblockA{\IEEEauthorrefmark{1}German Research Center for Artificial Intelligence (DFKI)\\Trippstadter Street 122,  Kaiserslautern, 67663 Germany\\
  }
\IEEEauthorblockA{\IEEEauthorrefmark{2}Technische Universit\"at  (RPTU) Kaiserslautern\\Building 11, Paul-Ehrlich Street, Kaiserslautern, 67663 Germany\\
 }
\thanks{This work was supported by the German Federal Ministry of Education and Research (BMBF) through \emph{Open6G-Hub} (Grant no.  \emph{16KISK003K}) and \emph{Open6GHub+} (Grant no.  \emph{16KIS2402K}) projects.}
}
\maketitle

\begin{abstract}
Traditional cellular networks struggle with poor quality of service (QoS) for cell-edge users, while cell-free (CF) systems offer uniform QoS but incur high roll-out costs due to acquiring numerous access point (AP) sites and deploying a large-scale optical fiber network to connect them. This paper proposes a cost-effective heterogeneous massive MIMO architecture that integrates centralized co-located antennas at a cell-center base station with distributed edge APs. By strategically splitting massive antennas between centralized and distributed nodes, the system maintains high user fairness comparable to CF systems but reduces infrastructure costs substantially, by minimizing the required number of AP sites and fronthaul connections. Numerical results demonstrate its superiority in balancing performance and costs compared to cellular and CF systems.
\end{abstract}

\section{Introduction}

In cellular networks, base stations (BSs) are deployed at the center of cells. However, users at the cell edge often suffer from poor quality of service (QoS) due to significant path loss and severe inter-cell interference. This creates a stark performance disparity between cell-center and cell-edge users \cite{Ref_jiang2024cost}. Due to its ability to providing uniform QoS for all users, cell-free massive MIMO (CFmMIMO) has recently drawn significant attention \cite{Ref_ngo2017cellfree}. It eliminates cell boundaries by leveraging distributed access points (APs) \cite{Ref_nayebi2017precoding}. Despite its potential, CFmMIMO deployment faces critical challenges \cite{Ref_jiang2024cost}. Acquiring hundreds of AP sites in densely populated regions is prohibitively expensive and sometimes infeasible due to public resistance (e.g., health concerns over RF radiation in residential areas), high leasing fees in private commercial spaces, and restrictions in sensitive zones such as government facilities. Additionally, the fronthaul network demands labor-intensive tasks like trenching and drilling, significantly increasing deployment time and costs.

It is desired to design an innovative network architecture to offer uniform QoS in a cost-efficient way. 
To address this, we proposed a hierarchical cell-free architecture as a cost-effective alternative to cell-free (CF) systems in our prior work \cite{Ref_jiang2024hierarchical}, and further extend its use in wide-band scenarios \cite{Ref_jiang2024heterogeneous}. Later, a magazine article \cite{Ref_zhang2024interdependent} discusses a similar hierarchical framework in a high-level overview. Meanwhile, recent studies have explored integrating CFmMIMO into legacy cellular networks, focusing on coexistence and cooperation mechanisms \cite{Ref_kim2022deployment, Ref_buzzi2024coexisting}.

In this paper, we advance the hierarchical philosophy into multi-cell cellular networks by introducing a technique, called heterogeneous massive MIMO (HmMIMO). Differing from \cite{Ref_kim2022deployment} and \cite{Ref_buzzi2024coexisting}—where cellular and CF are two independent systems—HmMIMO seamlessly integrates co-located and distributed antennas within a single system. It comprises:
\begin{itemize}
    \item A central base station (cBS) with a large antenna array at the center of each cell in a network of cells, functioning as the primary signal transceiver and central processing unit (CPU) for distributed nodes.
    \item Multiple edge access points (eAPs) deployed at cell edges and dead zones, interconnected with the cBS via fronthaul links.
\end{itemize}
By deriving closed-form spectral efficiency (SE) expressions for downlink/uplink transmissions and conducting numerical evaluation, this work demonstrates that HmMIMO maintains competitive user fairness (95\%-likely per-user SE) comparable to CFmMIMO while dramatically outperforming cellular systems at the cell edge, all achieved with significantly reduced infrastructure costs (encompassing the expenses of AP site acquisitions and fronthaul connections).

The paper is structured as follows: the next section presents the system model. 
Section III examines uplink (UL) transmission, deriving a closed-form expression for the achievable SE, while Section IV focuses on downlink (DL) transmission. Numerical results are provided in Section V, and the paper concludes in Section VI.

\begin{figure}[!t]
    \centering
    \subfloat[Cellular Massive MIMO: a BS with a large antenna array is positioned at the center of each cell within a network of cells.]{
    \includegraphics[width=0.325\textwidth]{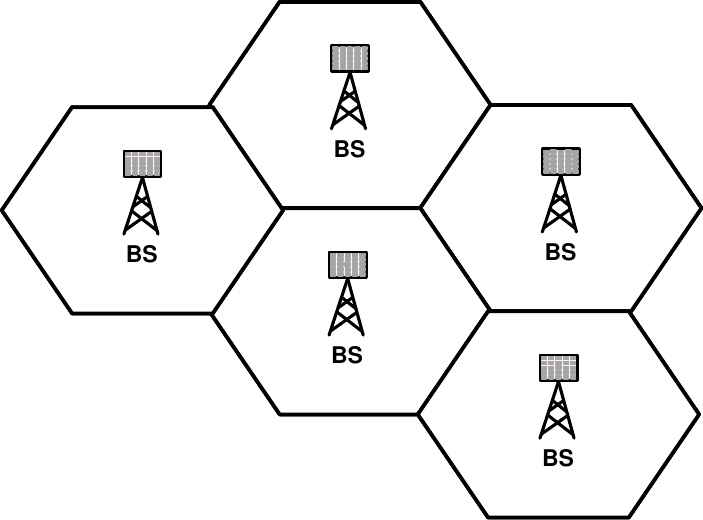} 
    } \\
    \subfloat[Cell-Free Massive MIMO: a massive number of distributed APs simultaneously serve users over an intended coverage area, coordinated by a CPU through a large-scale fronthaul network.]{
    \includegraphics[width=0.325\textwidth]{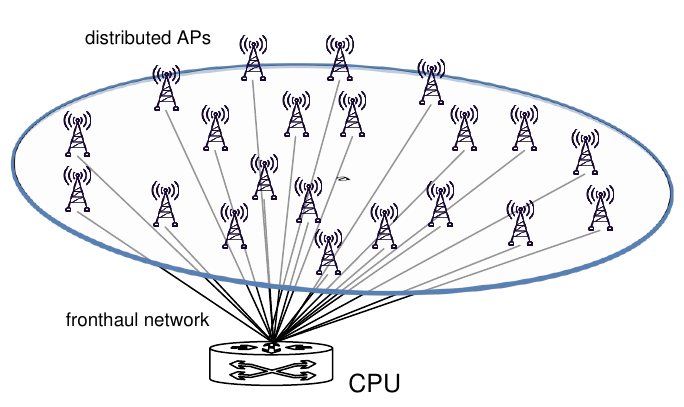} 
    }    \\
    \subfloat[heterogeneous Massive MIMO: each cell is covered by a cBS equipped with a large antenna array, along with some eAPs, which are connected to the cBS. ]{ \label{fig:heteromimo}
    \includegraphics[width=0.4\textwidth]{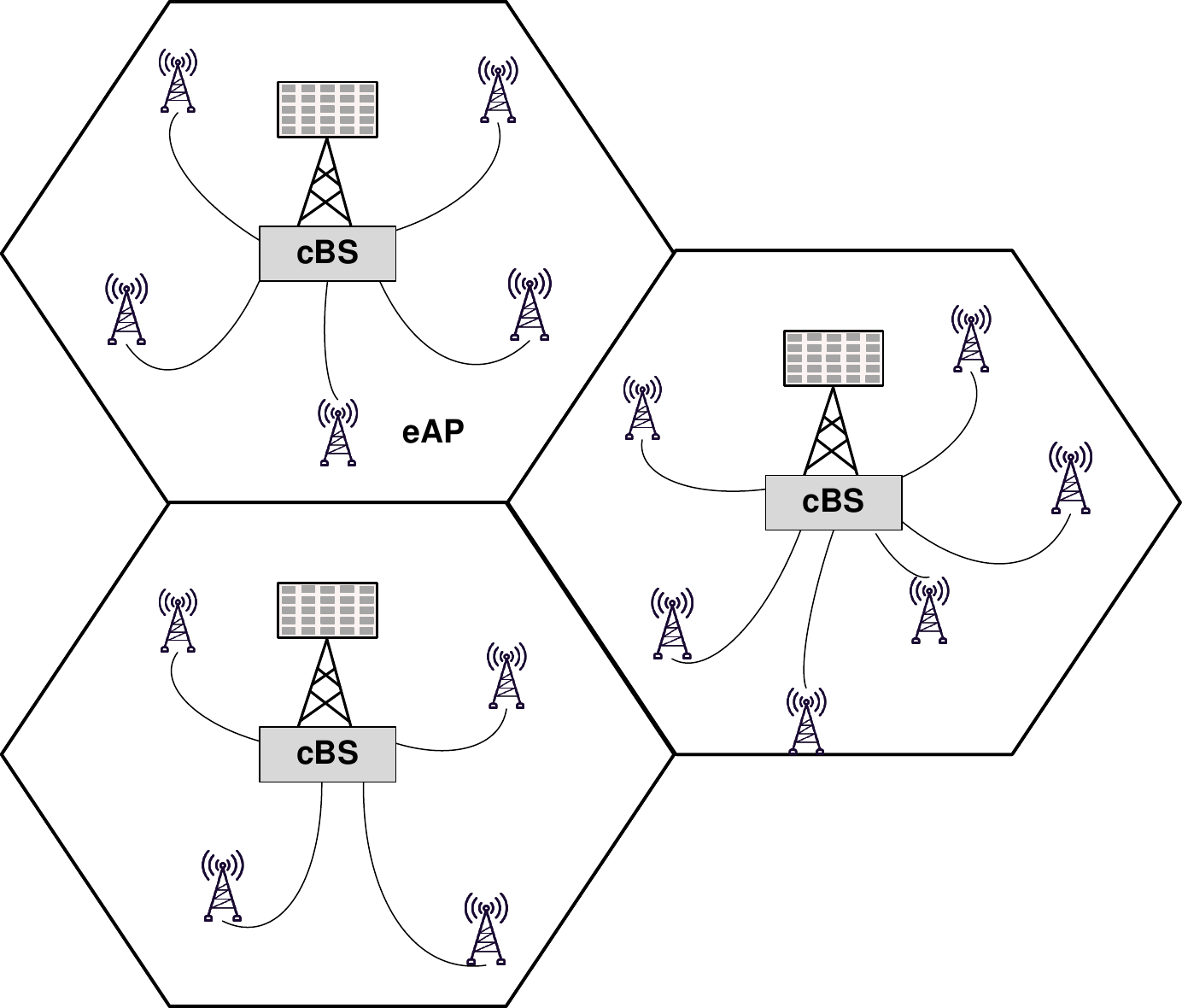} 
    }
    \caption{A comparative illustration of cellular, cell-free, and heterogeneous massive MIMO networks.  }
    \label{fig:SystemModel}
\end{figure}

\section{System Model}

This paper proposes a novel paradigm for cellular networking, as illustrated in \figurename \ref{fig:heteromimo}. Similar to cellular massive MIMO (CmMIMO) \cite{Ref_marzetta2015massive}, the intended area is covered by a cluster of cells. Each cell contains a cBS equipped with a large antenna array, positioned at the cell center.  In addition, a number of distributed eAPs are deployed in each cell and connected to the cBS through a fronthaul network. The use of eAPs aims to improve service quality, particularly in cell-edge regions or dead spots where the cBS's signal is weak. The cBS functions as both the primary signal transceiver for the cell and as the CPU for the eAPs within this cell.  Compared to CFmMIMO, this paradigm reduces the implementation cost of fronthaul networks since only a subset of service antennas in HmMIMO require connectivity.

Assume the network has $C$ cells, where a typical cell $c\in \{1,\ldots,C\}$ consists of a cBS with an array of $N_b$ co-located antennas, $L_c$ eAPs each with $N_a$ antennas, and $K_c$ single-antenna users. In contrast to some prior CFmMIMO studies that consider only single-antenna APs \cite{Ref_ngo2017cellfree, Ref_nayebi2017precoding}, this paper generalizes the system to multi-antenna eAPs. The correlation between co-located elements within an antenna array must, however, be considered. Under the block fading model, each coherent block refers to a time-frequency interval of $\tau_c$ channel uses during which the channel response remains approximately constant.
The channel between UE $k\in \{1,\ldots,K_{c}\}$ in cell $c$ and eAP $l \in \{1,\ldots,L_{\iota}\}$ in cell $\iota$ is denoted by $\mathbf{h}_{ck}^{\iota l}\in \mathbb{C}^{N_a}$. Each coherence block applies an independent realization from a \textit{correlated} Rayleigh fading  $\mathbf{h}_{ck}^{\iota l} \sim \mathcal{CN}(\mathbf{0}, \mathbf{R}_{ck}^{\iota l} )$\footnote{The channel realization is generated as $\mathbf{h} = \mathbf{R}^{1/2} \mathbf{h}'$, where $\mathbf{h}'$ represents independent and identically distributed Rayleigh fading with zero-mean and unit-variance elements \cite{yu2004modeling}.}, where $\mathbf{R}_{ck}^{\iota l}\in \mathbb{C}^{N_a\times N_a}$ stands for the spatial correlation matrix \cite{kermoal2002stochastic}. Likewise, the channel between the cBS in cell $\iota$ and UE $k$ of cell $c$ is represented by $\mathbf{h}_{ck}^{\iota0} \sim \mathcal{CN}(\mathbf{0}, \mathbf{R}_{ck}^{\iota0})$, where $\mathbf{R}_{ck}^{\iota0}\in \mathbb{C}^{N_b\times N_b}$. 

To avoid the significant overhead of inserting downlink pilots, which scales with the number of service antennas, time-division duplexing (TDD) is generally used in massive MIMO. This operation mode divides each coherent block into three phases: UL training, UL data transmission, and DL data transmission. In the UL training, each user equipment (UE) simultaneously transmits its pilot sequence, which has a length of $\tau_p$ symbols. Using linear minimum mean-square error (MMSE) for channel estimation \cite{Ref_bjornson2020making}, the estimate $\hat{\mathbf{h}}_{ck}^{\iota l}$, for all $l \in \{0, 1,\ldots,L_{\iota}\}$ follows the distribution $ \mathcal{CN}(\mathbf{0}, p_u \mathbf{R}_{ck}^{\iota l} (\boldsymbol{\Gamma}_{ck}^{\iota l})^{-1}\mathbf{R}_{ck}^{\iota l} )$, 
where $p_u$ denotes the power constraint per UE, and
\begin{equation}
     \boldsymbol{\Gamma}_{ck}^{\iota l}=\tau_p \left(p_u\tau_p\mathbf{R}_{ck}^{\iota l} + \sigma_n^2\mathbf{I}_{N_a}\right).
\end{equation}
The estimation error $\tilde{\mathbf{h}}_{ck}^{\iota l}=\mathbf{h}_{ck}^{\iota l} - \hat{\mathbf{h}}_{ck}^{\iota l}$ follows the distribution $\mathcal{CN}(\mathbf{0}, \boldsymbol{\Theta }_{ck}^{\iota l} )$ with its covariance matrix of 
\begin{equation}
\boldsymbol{\Theta}_{ck}^{\iota l}=\mathbb{E}\left[ \tilde{ \mathbf{h}}_{ck}^{\iota l} (\tilde{\mathbf{h}}_{ck}^{\iota l})^H \right]=\mathbf{R}_{ck}^{\iota l} - p_u \mathbf{R}_{ck}^{\iota l} (\boldsymbol{\Gamma}_{ck}^{\iota l})^{-1}\mathbf{R}_{ck}^{\iota l}. 
\end{equation}

\begin{figure*}[!t]
    \centering
    \includegraphics[width=0.55\textwidth]{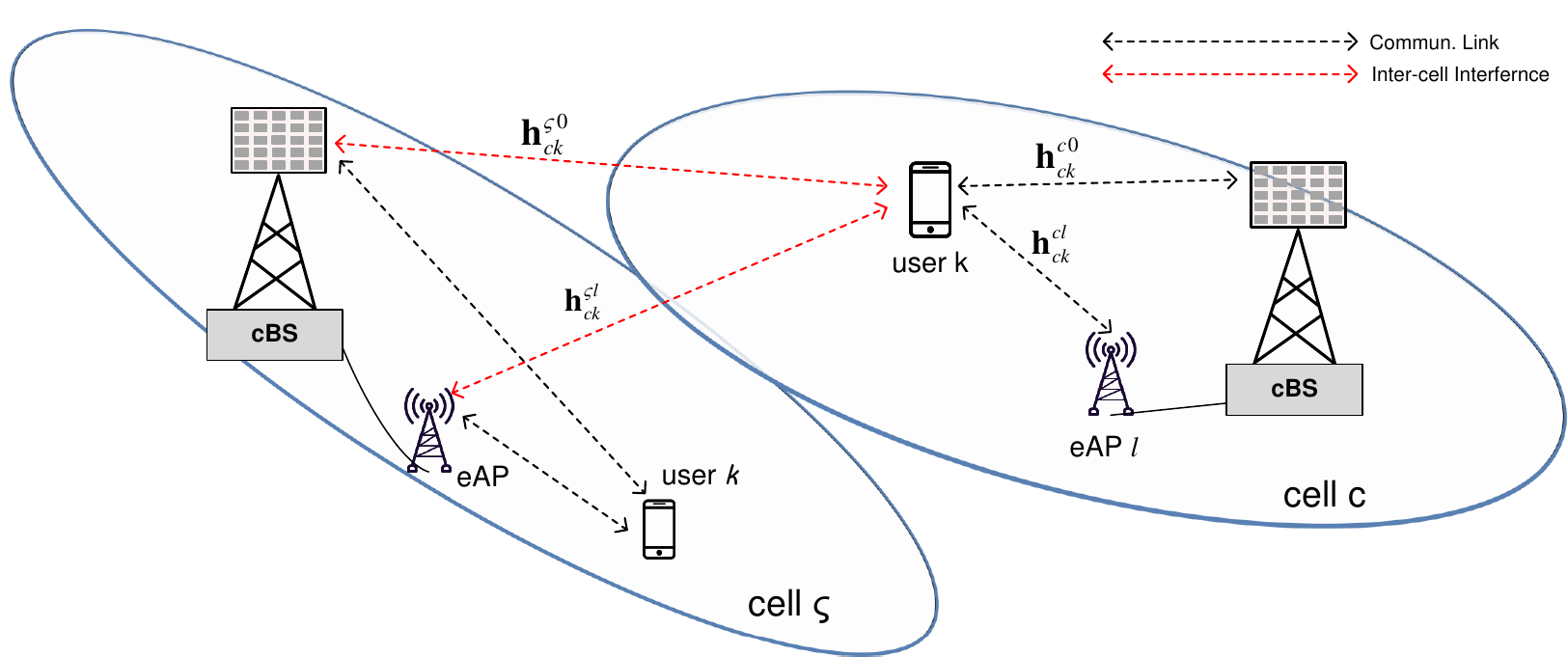}
    \caption{The system model of a toy example for the proposed heterogeneous massive MIMO with two adjacent cells.   }
    \label{diagram:SystemMode2}
\end{figure*}

\section{Uplink Data Transmission}
In the uplink, all UEs simultaneously send their signals, where UE $\kappa$ of cell $\iota$ sends $x_{\kappa\iota}$ with a power coefficient $0\leqslant \eta_{\kappa\iota} \leqslant 1$\footnote{In this paper, a \textit{general} user and its associated cell are denoted by $\kappa$ and $\iota$, respectively, while a \textit{particular} user and its cell, which are the focus of analysis and discussion, are represented by $k$ and $c$, respectively. }. The covariance matrix of the transmitted symbol vector $\mathbf{x}_{\iota}=[x_{\iota 1},\ldots,x_{\iota K_\iota}]^T$ satisfies $\mathbb{E}[\mathbf{x}_{\iota}\mathbf{x}_{\iota}^H]=\mathbf{I}_{K_{\iota}}$. Per channel use, a typical eAP $l$ in cell $c$ receives 
\begin{equation} \label{GS_uplink_RxsignalAP}
    \mathbf{y}_{cl} = \sqrt{p_u} \sum_{\iota=1}^C \sum_{\kappa=1}^{K_{\iota}} \sqrt{\eta_{\iota\kappa}} \mathbf{h}_{\iota\kappa}^{cl} x_{\iota\kappa} + \mathbf{n}_{cl}
\end{equation} 
with the receiver noise $\mathbf{n}_{cl}\sim \mathcal{CN}(\mathbf{0},\sigma^2_n\mathbf{I}_{N_a})$.
Meanwhile, the cBS of cell $c$ observes 
\begin{equation} \nonumber
    \mathbf{y}_{c0} = \sqrt{p_u} \sum_{\iota=1}^C \sum_{\kappa=1}^{K_{\iota}} \sqrt{\eta_{\iota\kappa}} \mathbf{h}_{\iota\kappa}^{c0} x_{\iota\kappa} + \mathbf{n}_{c0},
\end{equation} 
where $\mathbf{n}_{c0} \sim \mathcal{CN}(\mathbf{0},\sigma^2_n\mathbf{I}_{N_b})$.

Each eAP forwards its received signals to the cBS through fronthaul connections. As a result, the cBS collects $\mathbf{y}_c = [\mathbf{y}_{c0}^T,\mathbf{y}_{c1}^T\ldots,\mathbf{y}_{cL_c}^T]^T$,  which can be given by
\begin{equation} \label{eqnUPLINKmodel} \nonumber
    \mathbf{y}_c = \sqrt{p_u} \sum_{\iota=1}^C \sum_{\kappa=1}^{K_{\iota}} \sqrt{\eta_{\iota\kappa}} \mathbf{h}_{\iota\kappa}^{c} x_{\iota\kappa} + \mathbf{n}_{c},
\end{equation}
where $\mathbf{n}_{c} = [\mathbf{n}_{c0}^T,\mathbf{n}_{c1}^T\ldots,\mathbf{n}_{cL_c}^T]^T$, and $\mathbf{h}_{\iota \kappa}^{c}=\left[(\mathbf{h}_{\iota \kappa}^{c0})^T,(\mathbf{h}_{\iota \kappa}^{c1})^T\ldots,(\mathbf{h}_{\iota \kappa}^{cL_c})^T\right]^T$ denotes the complete set of channel coefficients between all service antennas of cell $c$ and a general user $\kappa$ in cell $\iota$. The spatial correlation matrix is represented by
\begin{align} \nonumber
    \mathbf{R}_{\iota \kappa}^c & =\mathbb{E}[\mathbf{h}_{\iota \kappa}^{c}(\mathbf{h}_{\iota \kappa}^{c})^H] = \mathrm{diag} (\mathbf{R}_{\iota \kappa}^{c0},\mathbf{R}_{\iota \kappa}^{c1},\ldots,\mathbf{R}_{\iota \kappa}^{cL_c}).
\end{align} 
The estimation error $\tilde{\mathbf{h}}_{\iota \kappa}^c=\mathbf{h}_{\iota \kappa}^c - \hat{\mathbf{h}}_{\iota \kappa}^c$ has the covariance matrix of
\begin{align} \nonumber
    \boldsymbol{\Theta}_{\iota \kappa}^c & = \mathbb{E}[\tilde{\mathbf{h}}_{\iota \kappa}^{c}(\tilde{\mathbf{h}}_{\iota \kappa}^{c})^H]= \mathrm{diag} (\boldsymbol{\Theta}_{\iota \kappa}^{c0},\boldsymbol{\Theta}_{\iota \kappa}^{c1},\ldots,\boldsymbol{\Theta}_{\iota \kappa}^{cL_c}).
\end{align} 

To recover $x_{ck}$, the cBS of cell $c$ performs maximum-ratio combining, as in \cite{Ref_ngo2017cellfree}, by multiplexing $\mathbf{y}_c$ with the Hermitian transpose of the channel estimate $\hat{\mathbf{h}}_{ck}^c$, resulting in
\begin{align} \label{massiveMIMO:MFsoftestimateUL} \nonumber
    \hat{x}_{ck} &=\frac{1}{\sqrt{p_u}} (\hat{\mathbf{h}}_{ck}^c)^H \mathbf{y}_c \\ \nonumber
    &=  \sum_{\iota=1}^C \sum_{\kappa=1}^{K_{\iota}} \sqrt{\eta_{\iota\kappa}} (\hat{\mathbf{h}}_{ck}^c)^H \mathbf{h}_{\iota\kappa}^{c} x_{\iota\kappa} + (\hat{\mathbf{h}}_{ck}^c)^H \frac{\mathbf{n}_{c} }{\sqrt{p_u}}
    \\ \nonumber
    &=\underbrace{ \sqrt{ \eta_{ck}} \|\hat{\mathbf{h}}_{ck}^c\|^2  x_{ck} }_{desired\:signal} + \underbrace{ \sqrt{ \eta_{ck}} (\hat{\mathbf{h}}_{ck}^c)^H \tilde{\mathbf{h}}_{ck}^c  x_{ck} }_{\mathcal{I}_1:\:channel\:estimation\:error} \\&+\underbrace{\sum_{{\kappa}=1,{\kappa}\neq k}^{K_c} \sqrt{\eta_{c\kappa}} (\hat{\mathbf{h}}_{ck}^c)^H \mathbf{h}_{c\kappa}^c  x_{c\kappa}}_{\mathcal{I}_2:\:intra-cell\:interference}\\ \nonumber
    &+ \underbrace{  \sum_{\iota=1,\iota\neq c}^C \sum_{\kappa=1}^{K_{\iota}} \sqrt{\eta_{\iota\kappa}} (\hat{\mathbf{h}}_{ck}^c)^H \mathbf{h}_{\iota\kappa}^{c} x_{\iota\kappa}}_{\mathcal{I}_3:\:inter-cell\:interference} +\underbrace{\frac{(\hat{\mathbf{h}}_{ck}^c)^H\mathbf{n}_c}{\sqrt{p_u}}}_{\mathcal{I}_4}.
\end{align} 
Therefore, the UL achievable SE of user $k$ in cell $c$ is
\begin{equation}
    C_{ck}^{UL}=\left(1-\frac{\tau_p}{\tau_u}\right)\mathbb{E} \Bigl[\log_2(1+\gamma_{ck}^{UL}) \Bigr],
\end{equation}
where the instantaneous effective signal-to-interference-plus-noise ratio (SINR) is given by
\begin{align} \label{eQn_ULspectraleff_ck}
    &\gamma_{ck}^{UL} = \\ \nonumber
    &\frac{ \eta_{ck} \|  \hat{\mathbf{h}}_{ck}^c \|^4  }{  (\hat{\mathbf{h}}_{ck}^c)^H \left( \begin{aligned}   
     & \sum\limits_{{\kappa}=1,{\kappa}\neq k}^{K_c} \eta_{c\kappa}\hat{\mathbf{h}}_{c\kappa}^c (\hat{\mathbf{h}}_{c\kappa}^c)^H + \sum\limits_{{\kappa}=1}^{K_c} \eta_{c\kappa} \boldsymbol{\Theta}_{c\kappa}^c \\   & \sum_{\iota=1,\iota\neq c}^C \sum_{\kappa=1}^{K_{\iota}} \eta_{\iota\kappa}   \mathbf{R}_{\iota\kappa}^c  +\frac{\sigma_n^2}{p_u}\mathbf{I}_{M_c} \end{aligned} \right) \hat{\mathbf{h}}_{ck}^c }
\end{align}

\begin{IEEEproof}
The interference terms from $\mathcal{I}_1$ to $\mathcal{I}_4$ in \eqref{massiveMIMO:MFsoftestimateUL} are mutually uncorrelated due to the independence of data symbols/noise, namely $\mathbb{E}[x_{ck}^*x_{\iota\kappa}]=0$, for all $c\neq \iota$ or $k\neq \kappa$. This implies $ \mathbb{E}[|\sum_{i=1}^4\mathcal{I}_i|^2]=\sum_{i=1}^4 \mathbb{E}[|\mathcal{I}_i|^2]$.
Note that $\{\hat{\mathbf{h}}_{c\kappa}^c\}_{\kappa=1,\ldots,K_c}$ are \textit{deterministic} for the cBS of cell $c$, the variances of the first two interference terms are computed as
\begin{align}  \label{APPEQ1} \nonumber
    \mathbb{E}\left[|\mathcal{I}_1|^2\right] & =  \eta_{ck} (\hat{\mathbf{h}}_{ck}^c)^H \mathbb{E}\left[\tilde{\mathbf{h}}_{ck}^c(\tilde{\mathbf{h}}_{ck}^c)^H\right] \hat{\mathbf{h}}_{ck}^c\\
    & =  \eta_{ck} (\hat{\mathbf{h}}_{ck}^c)^H \boldsymbol{\Theta}_{ck}^c \hat{\mathbf{h}}_{ck}^c,
\end{align}
and
\begin{align} \nonumber
    &\mathbb{E}\left[|\mathcal{I}_2|^2\right]=\sum_{{\kappa}=1,{\kappa}\neq k}^{K_c} \eta_{c\kappa}\mathbb{E}\left[\left| (\hat{\mathbf{h}}_{ck}^c)^H \mathbf{h}_{c\kappa}^c  \right|^2\right]\\ \nonumber
     &=\sum_{{\kappa}=1,{\kappa}\neq k}^{K_c} \eta_{c\kappa}(\hat{\mathbf{h}}_{ck}^c)^H \mathbb{E}\left[  \mathbf{h}_{c\kappa}^c (\mathbf{h}_{c\kappa}^c)^H \right] \hat{\mathbf{h}}_{ck}^c \\ \nonumber
    &=\sum_{{\kappa}=1,{\kappa}\neq k}^{K_c} \eta_{c\kappa}(\hat{\mathbf{h}}_{ck}^c)^H\left(\hat{\mathbf{h}}_{c\kappa}^c (\hat{\mathbf{h}}_{c\kappa}^c)^H+ \mathbb{E}\left[ \tilde{\mathbf{h}}_{c\kappa}^c (\tilde{\mathbf{h}}_{c\kappa}^c)^H \right] \right) \hat{\mathbf{h}}_{ck}^c \\
   &=\sum_{{\kappa}=1,{\kappa}\neq k}^{K_c} \eta_{c\kappa}(\hat{\mathbf{h}}_{ck}^c)^H\left(\hat{\mathbf{h}}_{c\kappa}^c (\hat{\mathbf{h}}_{c\kappa}^c)^H+ \boldsymbol{\Theta}_{c\kappa}^c \right) \hat{\mathbf{h}}_{ck}^c.   
\end{align}
In contrast, the cBS of cell $c$ does not know $\{\hat{\mathbf{h}}_{\iota\kappa}^c\}$ because it lacks direct connections to the eAPs in other cells. Consequently, the calculation of the third term is different, i.e.,
\begin{align} \nonumber
    \mathbb{E}\left[|\mathcal{I}_{3}|^2\right]&=\sum_{\iota=1,\iota\neq c}^C \sum_{\kappa=1}^{K_{\iota}} \eta_{\iota\kappa}\mathbb{E}\left[\left| (\hat{\mathbf{h}}_{ck}^c)^H \mathbf{h}_{\iota\kappa}^c  \right|^2\right]\\ \nonumber
    &=\sum_{\iota=1,\iota\neq c}^C \sum_{\kappa=1}^{K_{\iota}} \eta_{\iota\kappa}(\hat{\mathbf{h}}_{ck}^c)^H \mathbb{E}\left[  \mathbf{h}_{\iota\kappa}^c (\mathbf{h}_{\iota\kappa}^c)^H \right] \hat{\mathbf{h}}_{ck}^c \\
   &=\sum_{\iota=1,\iota\neq c}^C \sum_{\kappa=1}^{K_{\iota}} \eta_{\iota\kappa}(\hat{\mathbf{h}}_{ck}^c)^H   \mathbf{R}_{\iota\kappa}^c  \hat{\mathbf{h}}_{ck}^c . 
\end{align}
Lastly, 
\begin{equation}
    \mathbb{E}\left[|\mathcal{I}_4|^2\right] = \frac{\sigma_n^2}{p_u}(\hat{\mathbf{h}}_{ck}^c)^H \mathbf{I}_{M_c}  \hat{\mathbf{h}}_{ck}^c,
\end{equation}
where $M_c=N_b+L_cN_a$ denotes the number of service antennas in cell $c$. Thus, we get \eqref{eQn_ULspectraleff_ck}.
\end{IEEEproof}

\section{Downlink Data Transmission}

The UL and DL channels are reciprocal within a coherence block, enabling the cBS to use the UL channel estimates to precode the DL data symbols. The symbols intended for $K_{\iota}$ users in cell ${\iota}$ are denoted by $\mathbf{u}_{\iota}=[u_{{\iota}1},\ldots,u_{{\iota}K_{\iota}}]^T$, where  $\mathbb{E}[\mathbf{u}_{\iota}\mathbf{u}_{\iota}^H]=\mathbf{I}_{K_{\iota}}$. Using conjugate beamforming (CBF) \cite{Ref_yang2013performance} to spatially multiplex data symbols, a typical eAP $l$ in cell ${\iota}$ transmits the DL signal
\begin{equation} \label{eQn:compositeTxSig_MR}
    \mathbf{s}_{{\iota}l} = \sum_{\kappa=1}^{K_{\iota}} \mathbf{D}_{{\iota}\kappa}^{{\iota}l} ( \hat{\mathbf{h}}_{{\iota}\kappa}^{{\iota}l})^* u_{{\iota}\kappa},
\end{equation}
where $\mathbf{D}_{{\iota}\kappa}^{{\iota}l}$ is a diagonal matrix with power-control coefficients.
Meanwhile, the cBS sends 
\begin{equation}
    \mathbf{s}_{{\iota}0} = \sum_{\kappa=1}^{K_{\iota}} \mathbf{D}_{{\iota}\kappa}^{{\iota}0} ( \hat{\mathbf{h}}_{{\iota}\kappa}^{{\iota}0})^* u_{{\iota}\kappa}.
\end{equation} 
Consequently, the $k^{th}$ user in cell $c$ has the observation of
\begin{align} \nonumber \label{eQn_downlinkModel}
    y_{ck} &= \sqrt{p_d} \sum_{\iota=1}^C (\mathbf{h}_{ck}^{\iota0})^T\mathbf{s}_{\iota0} + \sqrt{p_d} \sum_{\iota=1}^C \sum_{l=1}^{L_{\iota}} (\mathbf{h}_{ck}^{\iota l})^T\mathbf{s}_{\iota l}  +n_{ck}\\ \nonumber
&= \sqrt{p_d} \sum_{\iota=1}^C (\mathbf{h}_{ck}^{\iota0})^T\sum_{\kappa=1}^{K_{\iota}} \mathbf{D}_{\iota\kappa}^{\iota0} ( \hat{\mathbf{h}}_{\iota\kappa}^{\iota0})^* u_{\iota\kappa} \\ &+ \sqrt{p_d} \sum_{\iota=1}^C \sum_{l=1}^{L_{\iota}} (\mathbf{h}_{ck}^{\iota l})^T\sum_{\kappa=1}^{K_{\iota}} \mathbf{D}_{\iota\kappa}^{\iota l} ( \hat{\mathbf{h}}_{\iota\kappa}^{\iota l})^* u_{\iota\kappa}  +n_{ck},
\end{align} 
where $n_{ck}\sim \mathcal{CN}(0,\sigma^2_n)$.

This user does not know channel estimates due to the absence of downlink pilots \cite{ngo2017no}.  As a result, the user detects the signals based on channel statistics $ \{ \mathbb{E} [  (\hat{\mathbf{h}}_{ck}^{cl})^H \hat{\mathbf{h}}_{ck}^{cl}  ] \}_{l \in \{0, 1,\ldots,L_{\iota}\} }$. Decomposing \eqref{eQn_downlinkModel} into \eqref{eQn_DLGeneralSig} shown on the next page, the achievable SE of user $k$ in the downlink is given by
\begin{equation}
    C_{ck}^{DL}=\mathbb{E} \Bigl[\log_2(1+\gamma_{ck}^{DL}) \Bigr],
\end{equation}
where the instantaneous SINR equals \eqref{eQn:DLSINR}.

\begin{figure*}[!t]
\begin{align} \nonumber \label{eQn_DLGeneralSig}
    y_{ck} &=\underbrace{\left( \mathbb{E} \left[   (\mathbf{h}_{ck}^{c0})^T \mathbf{D}_{ck}^{c0} ( \hat{\mathbf{h}}_{ck}^{c0})^* \right] + \sum_{l=1}^{L_c} \mathbb{E} \left[ (\mathbf{h}_{ck}^{cl})^T \mathbf{D}_{ck}^{cl} ( \hat{\mathbf{h}}_{ck}^{cl})^*  \right] \right) u_{ck}}_{desired\:signal} + \underbrace{\left( \begin{aligned} & (\mathbf{h}_{ck}^{c0})^T \mathbf{D}_{ck}^{c0} ( \hat{\mathbf{h}}_{ck}^{c0})^*-\mathbb{E} \left[   (\mathbf{h}_{ck}^{c0})^T \mathbf{D}_{ck}^{c0} ( \hat{\mathbf{h}}_{ck}^{c0})^* \right] + \\ &  \sum_{l=1}^{L_c} \left((\mathbf{h}_{ck}^{cl})^T \mathbf{D}_{ck}^{cl} ( \hat{\mathbf{h}}_{ck}^{cl})^*-\mathbb{E} \left[ (\mathbf{h}_{ck}^{cl})^T \mathbf{D}_{ck}^{cl} ( \hat{\mathbf{h}}_{ck}^{cl})^*  \right]\right) \end{aligned}\right)u_{ck} }_{\mathcal{J}_1:\:channel\:uncertainty\:error} \\ & + \underbrace{ (\mathbf{h}_{ck}^{c0})^T\sum_{\kappa=1,\kappa\neq k}^{K_{c}} \mathbf{D}_{c\kappa}^{c0} ( \hat{\mathbf{h}}_{c\kappa}^{c0})^* u_{c\kappa}+ \sum_{l=1}^{L_{c}} (\mathbf{h}_{ck}^{cl})^T\sum_{\kappa=1,\kappa\neq k}^{K_{c}} \mathbf{D}_{c\kappa}^{cl} ( \hat{\mathbf{h}}_{c\kappa}^{cl})^* u_{c\kappa} }_{\mathcal{J}_2:\:intra-cell\:interference} \\ \nonumber &+ \underbrace{  \sum_{\iota=1,\iota\neq c}^C (\mathbf{h}_{ck}^{\iota0})^T\sum_{\kappa=1}^{K_{\iota}} \mathbf{D}_{\iota\kappa}^{\iota0} ( \hat{\mathbf{h}}_{\iota\kappa}^{\iota0})^* u_{\iota\kappa}   + \sum_{\iota=1,\iota\neq c}^C \sum_{l=1}^{L_{\iota}} (\mathbf{h}_{ck}^{\iota l})^T\sum_{\kappa=1}^{K_{\iota}} \mathbf{D}_{\iota\kappa}^{\iota l} ( \hat{\mathbf{h}}_{\iota\kappa}^{\iota l})^* u_{\iota\kappa} }_{\mathcal{J}_3:\:inter-cell\:interference} + \frac{n_{ck}}{\sqrt{p_d}}
\end{align} 
\begin{equation} \label{eQn:DLSINR}
    \gamma_{ck}^{DL} = \frac{  \left| p_u \tr\left( \mathbf{D}_{ck}^{c0} 
\mathbf{R}_{ck}^{c0} (\boldsymbol{\Gamma}_{ck}^{c0})^{-1}\mathbf{R}_{ck}^{c0} \right) + \sum \limits_{l=1}^{L_c} p_u \tr\left( \mathbf{D}_{ck}^{cl} 
\mathbf{R}_{ck}^{cl} (\boldsymbol{\Gamma}_{ck}^{cl})^{-1}\mathbf{R}_{ck}^{cl} \right) \right|^2  }{      
     \sum \limits_{\iota=1}^C  \sum \limits_{\kappa=1}^{K_{\iota}} p_u \tr \left ( (  \mathbf{D}_{\iota\kappa}^{\iota  0} )^2 \mathbf{R}_{ck}^{\iota  0} \mathbf{R}_{\iota\kappa}^{\iota  0} (\boldsymbol{\Gamma}_{\iota\kappa}^{\iota  0})^{-1}\mathbf{R}_{\iota\kappa}^{\iota  0}    \right) +
     \sum \limits_{\iota=1}^C  \sum \limits_{l=1}^{L_\iota} \sum \limits_{\kappa=1}^{K_{\iota}} p_u \tr \left ( (  \mathbf{D}_{\iota\kappa}^{\iota  l} )^2 \mathbf{R}_{ck}^{\iota  l} \mathbf{R}_{\iota\kappa}^{\iota  l} (\boldsymbol{\Gamma}_{\iota\kappa}^{\iota  l})^{-1}\mathbf{R}_{\iota\kappa}^{\iota  l}    \right)+ \frac{\sigma_n^2}{p_d}
      } 
\end{equation}
\end{figure*}

\begin{IEEEproof} First, we have
\begin{align} \label{eQn:expectedValuedforhkzero} \nonumber
    \mathbb{E} \left[   (\mathbf{h}_{ck}^{c0})^T \mathbf{D}_{ck}^{c0} ( \hat{\mathbf{h}}_{ck}^{c0})^* \right] &= \mathbb{E} \left[   ( \hat{\mathbf{h}}_{ck}^{c0}+\tilde{\mathbf{h}}_{ck}^{c0})^T \mathbf{D}_{ck}^{c0} ( \hat{\mathbf{h}}_{ck}^{c0})^* \right] \\
    &= \mathbb{E} \left[    (\hat{\mathbf{h}}_{ck}^{c0})^T \mathbf{D}_{ck}^{c0} ( \hat{\mathbf{h}}_{ck}^{c0})^* \right],
\end{align}
as channel estimates and their errors are independent, i.e., $\mathbb{E}[( \hat{\mathbf{h}}_{ck}^{c0})^H\tilde{\mathbf{h}}_{ck}^{c0} ] =0$.
Further, it follows that 
\begin{align} \label{Proof_expectedValueHk0}
    \mathbb{E} \left[    (\hat{\mathbf{h}}_{ck}^{c0})^T \mathbf{D}_{ck}^{c0} ( \hat{\mathbf{h}}_{ck}^{c0})^* \right] &= \tr \left( \mathbb{E} \left[    \mathbf{D}_{ck}^{c0}  \hat{\mathbf{h}}_{ck}^{c0}(\hat{\mathbf{h}}_{ck}^{c0})^H \right]  \right)
    \\ \nonumber &=p_u \tr\left( \mathbf{D}_{ck}^{c0} 
\mathbf{R}_{ck}^{c0} (\boldsymbol{\Gamma}_{ck}^{c0})^{-1}\mathbf{R}_{ck}^{c0} \right),
\end{align} where $\tr$ denotes the trace of a matrix,
because
\begin{equation}
    \mathbb{E} \left[    \hat{\mathbf{h}}_{ck}^{c0} ( \hat{\mathbf{h}}_{ck}^{c0})^H \right] = p_u \mathbf{R}_{ck}^{c0} (\boldsymbol{\Gamma}_{ck}^{c0})^{-1}\mathbf{R}_{ck}^{c0}.
\end{equation}
Likewise, 
\begin{equation} \label{eQn_drivationyyyy}
    \mathbb{E} \left[    (\mathbf{h}_{ck}^{cl})^T \mathbf{D}_{ck}^{cl} ( \hat{\mathbf{h}}_{ck}^{cl})^* \right] = p_u \tr\left( \mathbf{D}_{ck}^{cl} 
\mathbf{R}_{ck}^{cl} (\boldsymbol{\Gamma}_{ck}^{cl})^{-1}\mathbf{R}_{ck}^{cl} \right).
\end{equation}

Due to the independence of data symbols, namely $\mathbb{E}[u_{ck}^*u_{\iota\kappa}]=0$, for all $c\neq \iota$ or $k\neq \kappa$, the interference terms in \eqref{eQn_DLGeneralSig} are mutually uncorrelated. As a result, 
\begin{equation}
    \mathbb{E}\left[|\mathcal{J}_1+\mathcal{J}_2+\mathcal{J}_3|^2\right]= \mathbb{E}\left[|\mathcal{J}_1|^2\right] + \mathbb{E}\left[|\mathcal{J}_2|^2\right] +\mathbb{E}\left[|\mathcal{J}_3|^2\right].
\end{equation}
The variance of the first interference term is computed as
\begin{align}  \label{eQn_Jzero}
    & \mathbb{E}\left[|\mathcal{J}_1|^2\right] =\\ \nonumber
    &\sum_{l=0}^{L_c} \mathbb{E} \left[  \left| (\mathbf{h}_{ck}^{cl})^T \mathbf{D}_{ck}^{cl} ( \hat{\mathbf{h}}_{ck}^{cl})^*-\mathbb{E} \left[ (\mathbf{h}_{ck}^{cl})^T \mathbf{D}_{ck}^{cl} ( \hat{\mathbf{h}}_{ck}^{cl})^*  \right]\right|^2   \right] =\\ \nonumber
    & \sum_{l=0}^{L_c} \left (\mathbb{E} \left[  \left| (\mathbf{h}_{ck}^{cl})^T \mathbf{D}_{ck}^{cl} ( \hat{\mathbf{h}}_{ck}^{cl})^*\right|^2\right] - \mathbb{E}^2 \left[ (\mathbf{h}_{ck}^{cl})^T \mathbf{D}_{ck}^{cl} ( \hat{\mathbf{h}}_{ck}^{cl})^*  \right] \right ).
\end{align}
The first component of \eqref{eQn_Jzero} can be further derived as
\begin{align} \nonumber \label{eQN:expeResults} \nonumber
     &\mathbb{E} \left[  \left| (\mathbf{h}_{ck}^{cl})^T \mathbf{D}_{ck}^{cl} ( \hat{\mathbf{h}}_{ck}^{cl})^*\right|^2\right]  \\  \nonumber
     &= \mathbb{E}[ (\mathbf{h}_{ck}^{cl})^T \mathbf{D}_{ck}^{cl} ( \hat{\mathbf{h}}_{ck}^{cl})^* ( \hat{\mathbf{h}}_{ck}^{cl})^T  \mathbf{D}_{ck}^{cl} (\mathbf{h}_{ck}^{cl})^* ] \\ \nonumber
      &= \mathbb{E}[ ( \hat{\mathbf{h}}_{ck}^{cl})^T \mathbf{D}_{ck}^{cl} ( \hat{\mathbf{h}}_{ck}^{cl})^* ( \hat{\mathbf{h}}_{ck}^{cl})^T  \mathbf{D}_{ck}^{cl} (\hat{\mathbf{h}}_{ck}^{cl})^* ]\\ 
      &+ \mathbb{E}[ (\tilde{\mathbf{h}}_{ck}^{cl})^T \mathbf{D}_{ck}^{cl} ( \hat{\mathbf{h}}_{ck}^{cl})^* ( \hat{\mathbf{h}}_{ck}^{cl})^T  \mathbf{D}_{ck}^{cl} (\tilde{\mathbf{h}}_{ck}^{cl})^* ],
\end{align}
where
\begin{align}  \label{eQn_FirstIntTerm} \nonumber
    &\mathbb{E}[ ( \hat{\mathbf{h}}_{ck}^{cl})^T \mathbf{D}_{ck}^{cl} ( \hat{\mathbf{h}}_{ck}^{cl})^* ( \hat{\mathbf{h}}_{ck}^{cl})^T  \mathbf{D}_{ck}^{cl} (\hat{\mathbf{h}}_{ck}^{cl})^* ] \\ \nonumber
    & = \mathbb{E}[ |(\hat{\mathbf{h}}_{ck}^{cl})^H \mathbf{D}_{ck}^{cl} \hat{\mathbf{h}}_{ck}^{cl} |^2 ]  \\ \nonumber
    &= p_u^2 \left| \tr(    \mathbf{D}_{ck}^{cl} \mathbf{R}_{ck}^{cl}  (\boldsymbol{\Gamma}_{ck}^{cl})^{-1}\mathbf{R}_{ck}^{cl} )  \right|^2\\  & 
    + p_u\tr(    (\mathbf{D}_{ck}^{cl})^2 ( \mathbf{R}_{ck}^{cl} - \boldsymbol{\Theta}_{ck}^{cl} )  \mathbf{R}_{ck}^{cl} (\boldsymbol{\Gamma}_{ck}^{cl})^{-1} \mathbf{R}_{ck}^{cl} ),
\end{align}
and
\begin{align}   \label{eQn_secondIntTerm} \nonumber
    &\mathbb{E}[ (\tilde{\mathbf{h}}_{ck}^{cl})^T \mathbf{D}_{ck}^{cl} ( \hat{\mathbf{h}}_{ck}^{cl})^* ( \hat{\mathbf{h}}_{ck}^{cl})^T  \mathbf{D}_{ck}^{cl} (\tilde{\mathbf{h}}_{ck}^{cl})^* ]\\ \nonumber
    &= \tr \left ((\mathbf{D}_{ck}^{cl})^2 \mathbb{E}[ \hat{\mathbf{h}}_{ck}^{cl} (\hat{\mathbf{h}}_{ck}^{cl})^H ] \mathbb{E}[\tilde{ \mathbf{h}}_{ck}^{cl} (\tilde{\mathbf{h}}_{ck}^{cl})^H  ] \right) \\
     &= p_u \tr \left ((\mathbf{D}_{ck}^{cl})^2 \boldsymbol{\Theta}_{ck}^{cl} \mathbf{R}_{ck}^{cl} (\boldsymbol{\Gamma}_{ck}^{cl})^{-1}\mathbf{R}_{ck}^{cl}    \right).
\end{align} 
Substituting \eqref{eQn_FirstIntTerm} and \eqref{eQn_secondIntTerm} into \eqref{eQN:expeResults}, yield
\begin{align} \nonumber \label{eQn_xxxxx}
    \mathbb{E} & \left[  \left|  (\mathbf{h}_{ck}^{cl})^T \mathbf{D}_{ck}^{cl} ( \hat{\mathbf{h}}_{ck}^{cl})^*\right|^2\right] =  p_u^2 \left| \tr(    \mathbf{D}_{ck}^{cl} \mathbf{R}_{ck}^{cl}  (\boldsymbol{\Gamma}_{ck}^{cl})^{-1}\mathbf{R}_{ck}^{cl} )  \right|^2 \\ 
    &+ p_u\tr(    (\mathbf{D}_{ck}^{cl})^2  \mathbf{R}_{ck}^{cl}   \mathbf{R}_{ck}^{cl} (\boldsymbol{\Gamma}_{ck}^{cl})^{-1} \mathbf{R}_{ck}^{cl} ).
\end{align}
Finally, we apply \eqref{eQn_xxxxx} and \eqref{eQn_drivationyyyy} in \eqref{eQn_Jzero} to obtain
\begin{align}  \label{eQn_JzeroFinal}
    \mathbb{E}\left[|\mathcal{J}_1|^2\right] &= \sum_{l=0}^{L_c} p_u\tr\left(    (\mathbf{D}_{ck}^{cl})^2  \mathbf{R}_{ck}^{cl}   \mathbf{R}_{ck}^{cl} (\boldsymbol{\Gamma}_{ck}^{cl})^{-1} \mathbf{R}_{ck}^{cl} \right).
\end{align}
Next, the second interference term follows 
\begin{align} \nonumber \label{eQn_Jtwo}
    &\mathbb{E}\left[|\mathcal{J}_2|^2\right] = \sum_{l=0}^{L_c} \sum_{\kappa=1,\kappa\neq k}^{K_{c}} \mathbb{E}[ (\mathbf{h}_{ck}^{cl})^T \mathbf{D}_{c\kappa}^{cl} (\hat{\mathbf{h}}_{c\kappa}^{cl})^*  (\mathbf{h}_{ck}^{cl})^H \mathbf{D}_{c\kappa}^{cl} \hat{\mathbf{h}}_{c\kappa}^{cl} ] \\ \nonumber
    & = \sum_{l=0}^{L_c} \sum_{\kappa=1,\kappa\neq k}^{K_{c}} \tr \left ( ( \mathbf{D}_{c\kappa}^{cl}  )^2 \mathbb{E}[ \mathbf{h}_{ck}^{cl}(\mathbf{h}_{ck}^{cl})^H ] \mathbb{E}[ \hat{\mathbf{h}}_{c\kappa}^{cl} (\hat{\mathbf{h}}_{c\kappa}^{cl})^H  ] \right) \\
     &= \sum_{l=0}^{L_c} \sum_{\kappa=1,\kappa\neq k}^{K_{c}} p_u \tr \left ( ( \mathbf{D}_{c\kappa}^{cl}  )^2 \mathbf{R}_{ck}^{cl} \mathbf{R}_{c\kappa}^{cl} (\boldsymbol{\Gamma}_{c\kappa}^{cl})^{-1}\mathbf{R}_{c\kappa}^{cl}    \right).
\end{align}
Likewise, the variance of $\mathcal{J}_3$ equals
\begin{align}  \label{eQn_Jthree}
    &\mathbb{E}\left[|\mathcal{J}_3|^2\right]= \\ \nonumber
    & \sum_{\iota=1,\iota\neq c}^C  \sum_{l=0}^{L_\iota} \sum_{\kappa=1}^{K_{\iota}} \mathbb{E}[ (\mathbf{h}_{ck}^{\iota  l})^T \mathbf{D}_{\iota\kappa}^{\iota  l} ( \hat{\mathbf{h}}_{\iota\kappa}^{\iota  l})^* (\mathbf{h}_{ck}^{\iota  l})^H \mathbf{D}_{\iota\kappa}^{\iota  l}  \hat{\mathbf{h}}_{\iota\kappa}^{\iota  l} ] \\ \nonumber
    & =  \sum_{\iota=1,\iota\neq c}^C  \sum_{l=0}^{L_\iota} \sum_{\kappa=1}^{K_{\iota}} \tr \left ( (  \mathbf{D}_{\iota\kappa}^{\iota  l} )^2 \mathbb{E}[ \mathbf{h}_{ck}^{\iota  l} (\mathbf{h}_{ck}^{\iota  l})^H  ] \mathbb{E}[  \hat{\mathbf{h}}_{\iota\kappa}^{\iota  l} ( \hat{\mathbf{h}}_{\iota\kappa}^{\iota  l})^H ] \right) \\ \nonumber
     &= \sum_{\iota=1,\iota\neq c}^C  \sum_{l=0}^{L_\iota} \sum_{\kappa=1}^{K_{\iota}} p_u \tr \left ( (  \mathbf{D}_{\iota\kappa}^{\iota  l} )^2 \mathbf{R}_{ck}^{\iota  l} \mathbf{R}_{\iota\kappa}^{\iota  l} (\boldsymbol{\Gamma}_{\iota\kappa}^{\iota  l})^{-1}\mathbf{R}_{\iota\kappa}^{\iota  l}    \right).
\end{align}
Utilizing \eqref{eQn_JzeroFinal}, \eqref{eQn_Jtwo}, and \eqref{eQn_Jthree} gets \eqref{eQn:DLSINR}.
\end{IEEEproof}

\section{Numerical Results}

We evaluate the performance of HmMIMO against CFmMIMO and cellular mMIMO in terms of SE and user fairness. Our simulations apply a square coverage area of $1 \mathrm{km} \times 1 \mathrm{km}$, partitioned into four $500 \mathrm{m} \times 500 \mathrm{m}$ square cells for the HmMIMO and cellular setup. For a fair comparison, all three paradigms use $512$ antennas to serve $32$ users. In CFmMIMO, to be specific, $128$ APs, each with four antennas, are randomly distributed throughout the area. In the cellular system, each cell has a BS equipped with $128$ co-located antennas, serving $8$ users. Two HmMIMO configurations, i.e., a quarter or half of the antennas of a cell are aggregated to its cBS, i.e., $N_b = 32$ or 64, while each cell has $24$ or $16$ four-antenna eAPs. Across $10^5$ simulation epochs, eAPs/APs and users are randomly located to capture the ergodic performance.

The path loss of channels is determined by the COST-Hata model, given in \cite{Ref_ngo2017cellfree}. The shadowing loss is modeled as a log-normal variable $\mathcal{X} \sim \mathcal{N}(0, 8^2)$.
The white noise power density is \( -174\ \mathrm{dBm/Hz} \), with a noise figure of \( 9\ \mathrm{dB} \), and the signal bandwidth equals \( 5\ \mathrm{MHz} \). All arrays across BSs/cBSs and APs/eAPs are configured as uniform linear arrays with half-wavelength spacing between elements. Spatial correlation is represented using the Gaussian local scattering model with an angular standard deviation of \( 15^\circ \) \cite{Ref_bjornson2020making}. Each coherence block contains \( \tau_c = 200 \) channel uses, with \( \tau_p = 8 \) allocated for pilot sequences. Two power-control strategies: 1) full (uplink) and equal (downlink) power transmission; and 2) max-min power control, are applied.

\begin{figure}[!t]
    \centering
    \subfloat[]{
    \includegraphics[width=0.35\textwidth]{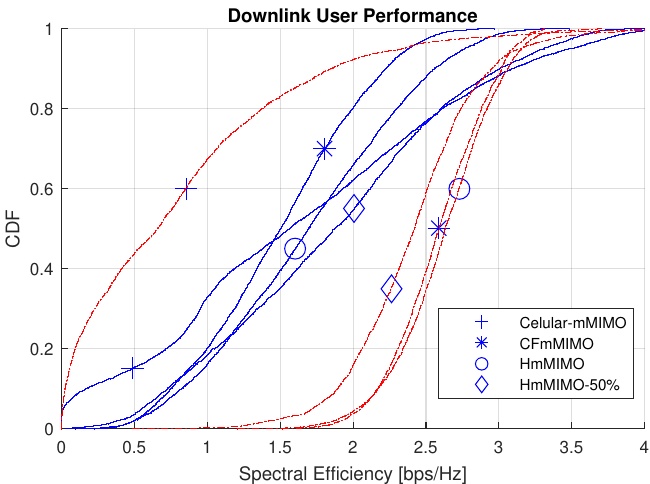} 
    \label{fig:siml}
    }  
    \\
    \subfloat[]{
    \includegraphics[width=0.35\textwidth]{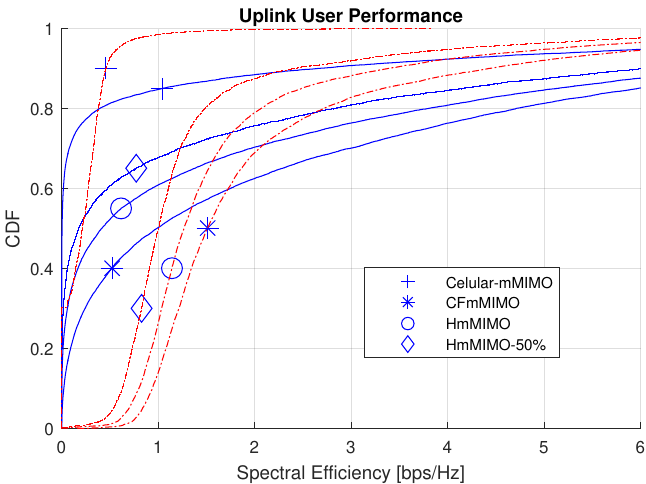} 
    \label{fig:sim2}  }    
    \caption{Performance comparison of cellular, cell-free, and heterogeneous massive MIMO networks in per-user SE.  }
    \label{fig:results}
\end{figure}

We compare downlink performance of HmMIMO, CFmMIMO, and cellular mMIMO networks via cumulative distribution functions (CDFs) of per-user SE (see \figurename~\ref{fig:siml}). The $95\%\text{-likely}$ user rate, marking cell-edge performance, is the $5^{th}$ percentile of these CDFs. Configurations are identified by distinct markers: solid lines denote equal power allocation, while dash-dotted lines with the same markers represent scenarios using max-min power control. HmMIMO (1/4 antenna aggregation) achieves $0.65~\mathrm{bps/Hz}$, topping CFmMIMO ($0.61~\mathrm{bps/Hz}$) and vastly outpacing cellular networks ($0.013~\mathrm{bps/Hz}$). Max-min control boosts distributed systems: CFmMIMO-maxmin hits $2.05~\mathrm{bps/Hz}$, HmMIMO-maxmin reaches $2.03~\mathrm{bps/Hz}$, but cellular systems drop to $0.006~\mathrm{bps/Hz}$ (from $0.013~\mathrm{bps/Hz}$). This occurs as cellular max-min allocates most power to the weakest cell-edge user, slashing power for cell-center users inefficiently. Higher centralization in HmMIMO-1/2 (half antennas aggregated) lowers the $95\%\text{-likely}$ rate to $0.55~\mathrm{bps/Hz}$ (vs. $0.65~\mathrm{bps/Hz}$), but lowering the number of wireless AP sites and cabling by $\sim\mathbf{50\%}$, thus saving remarkably infrastructure costs.

The results for uplink scenarios are illustrated in \figurename~\ref{fig:sim2}. In this figure, solid lines denote full power transmission by all users, while dash-dotted lines represent scenarios using max-min power control. Cellular networks demonstrate poor performance, underscoring their inadequacy for users at the cell edges. Conversely, CFmMIMO and HmMIMO show significant advantages, especially when employing max-min power control, with $95\%\text{-likely}$ user rates reaching 0.83~bps/Hz and 0.70~bps/Hz, respectively. Although HmMIMO experiences a slight degradation relative to CFmMIMO, it nonetheless significantly surpasses the cellular network. Notably, HmMIMO maintains a high level of fairness while decreasing the fronthaul scale by about $\mathbf{25\%}$. The HmMIMO-1/2 variant sacrifices some additional performance to attain even higher cost efficiency, resulting in a further decrease in $95\%\text{-likely}$ rate of 0.54~bps/Hz, yet it accomplishes a remarkable $\sim \mathbf{50\%}$ decrease in the number of wireless AP sites and cabling, making it a cost-efficient option.

\section{Conclusion}
This paper proposed a heterogeneous massive MIMO approach for cellular networks, which seamlessly integrates co-located and distributed antennas. Each cell has a central base station with a large antenna array at its center, aided by multiple edge access points positioned at the cell edges or in dead spots. The signal transmission and reception of eAPs are coordinated by the cBS through the fronthaul network. Numerical evaluation demonstrates that this paradigm substantially improves worst-case user rates comparable to those of traditional cellular networks, while significantly reducing the infrastructure costs of CFmMIMO.

\bibliographystyle{IEEEtran}
\bibliography{IEEEabrv,Ref_COML}

\end{document}